\documentclass[prb,twocolumn,showpacs,superscriptaddress,floatfix]{revtex4}

\usepackage{graphicx}
\usepackage[centertags]{amsmath}

\newcommand{\cm}{\ensuremath{\,\mbox{cm}^{-1}}}
\newcommand{\K}{\ensuremath{\,\mbox{K}}}
\begin{document}

\title{Antiferrodistortive phase transition in EuTiO$_{3}$}

\author{V.~Goian}
\affiliation{Institute of Physics ASCR, Na Slovance~2, 182 21 Prague~8, Czech
Republic}
\author{ S.~Kamba}\email{kamba@fzu.cz} \affiliation{Institute of Physics ASCR,
Na Slovance~2, 182 21 Prague~8, Czech Republic}
\author{ O.~Pacherov\'{a}} \affiliation{Institute of Physics ASCR,
Na Slovance~2, 182 21 Prague~8, Czech Republic}
\author{ J.~Drahokoupil} \affiliation{Institute of Physics ASCR,
Na Slovance~2, 182 21 Prague~8, Czech Republic}
\author{ L.~Palatinus} \affiliation{Institute of Physics ASCR,
Na Slovance~2, 182 21 Prague~8, Czech Republic}
\author{ M.~Du\v{s}ek} \affiliation{Institute of Physics ASCR,
Na Slovance~2, 182 21 Prague~8, Czech Republic}
\author{ J.~Rohl\'{i}\v{c}ek} \affiliation{Institute of Physics ASCR,
Na Slovance~2, 182 21 Prague~8, Czech Republic}
\author{ M. Savinov} \affiliation{Institute of Physics ASCR,
Na Slovance~2, 182 21 Prague~8, Czech Republic}
\author{F. Laufek}
\affiliation{Czech Geological Survey, Geologick\'{a}~6, 152 00 Prague~5, Czech Republic}
\author{W.~Schranz}
\affiliation{Faculty of Physics, University of Vienna,
Boltzmanngasse~5, A-1090 Wien, Austria}
\author{A.~Fuith}
\affiliation{Faculty of Physics, University of Vienna,
Boltzmanngasse~5, A-1090 Wien, Austria}
\author{M.~Kachl\'{i}k}
\affiliation{Department of Ceramics and Polymers, Brno University of Technology,
Technick\'{a}~2, 616 69 Brno, Czech Republic}
\affiliation{CEITEC BUT, Brno University of
Technology, Technick\'{a} 10, 616 00 Brno, Czech Republic}
\author{K.~Maca}
\affiliation{Department of Ceramics and Polymers, Brno University of Technology,
Technick\'{a}~2, 616 69 Brno, Czech Republic}
\affiliation{CEITEC BUT, Brno University of
Technology, Technick\'{a} 10, 616 00 Brno, Czech Republic}
\author{A. Shkabko}
\affiliation{Swiss Federal Laboratories for Materials Science and Technology,
Überlandstrasse 129, 8600 D\"{u}bendorf, Switzerland}
\author{L. Sagarna}
\affiliation{Swiss Federal Laboratories for Materials Science and Technology,
Überlandstrasse 129, 8600 D\"{u}bendorf, Switzerland}
\author{A. Weidenkaff}
\affiliation{Swiss Federal Laboratories for Materials Science and Technology,
Überlandstrasse 129, 8600 D\"{u}bendorf, Switzerland}
\author{A.A.~Belik}
\affiliation{International Center for Materials Nanoarchitectonics (WPI-MANA), National
Institute for Materials Science (NIMS), 1-1 Namiki, Tsukuba, Ibaraki 305-0044, Japan}

\date{\today}

\pacs{75.80.+q; 78.30.-j; 63.20.-e}

\begin{abstract}

X-ray diffraction, dynamical mechanical analysis and infrared reflectivity studies
revealed an antiferrodistortive phase transition in EuTiO$_{3}$ ceramics. Near 300\K\,
the perovskite structure changes from cubic $Pm\bar{3}m$ to tetragonal $I4/mcm$ due to
antiphase tilting of oxygen octahedra along the \textbf{c} axis ($a^{0}a^{0}c^{-}$ in
Glazer notation). The phase transition is analogous to SrTiO$_{3}$. However, some
ceramics as well as single crystals of EuTiO$_{3}$ show different infrared reflectivity
spectra bringing evidence of a different crystal structure. In such samples electron
diffraction revealed an incommensurate tetragonal structure with modulation wavevector
\textbf{q} $\simeq$ 0.38 \textbf{a}$^{*}$. Extra phonons in samples with modulated
structure are activated in the IR spectra due to folding of the Brillouin zone. We
propose that defects like Eu$^{3+}$ and oxygen vacancies strongly influence the
temperature of the phase transition to antiferrodistortive phase as well as the tendency
to incommensurate modulation in EuTiO$_{3}$.

\end{abstract}

\maketitle

\section{Introduction}

EuTiO$_{3}$ is a frequently investigated material in the last decade thanks to Katsufuji
and Takagi,\cite{katsufuji01} who discovered a strong magnetodielectric effect in
antiferromagnetic (AFM) G-type phase\cite{guire66} of this material below T$_{N}$ =
5.3\K. 7\% change of permittivity with magnetic field was found at 2\K. Linear
magnetoelectric coupling is forbidden in EuTiO$_{3}$ due to centrosymmetric structure of
this material. Quadratic coupling was not detected, but a strong third-order
(bielectrobimagnetic $\sim E^{2}H^{2}$) magnetoelectric coupling was observed in
Ref.\cite{shvartsman10}. Dielectric permittivity $\varepsilon$' exhibits a typical
incipient ferroelectric behavior: $\varepsilon$' increases on cooling and saturates below
50\K.\cite{katsufuji01} This temperature behavior was explained by an optical phonon
softening on cooling and by saturation of its frequency at low
temperatures.\cite{kamba07,goian09} Both temperature dependences of $\varepsilon$' and
soft phonon frequency follow the Barrett formula,\cite{kamba07,goian09} which takes into
account quantum fluctuations at low temperatures. $\varepsilon$' drops down below T$_{N}$
by several percents due to a strong spin-phonon coupling.\cite{katsufuji01} In AFM phase
the temperature and magnetic field dependence of $\varepsilon$' is caused by the response
of the lowest-frequency phonon to the magnetic order and/or magnetic field.\cite{kamba12}

Fennie and Rabe\cite{fennie06} suggested to use the spin-phonon coupling and a biaxial
strain in the thin EuTiO$_{3}$ films for induction of ferroelectric and ferromagnetic
order, although the bulk EuTiO$_{3}$ is quantum paraelectric and antiferromagnetic.
Recently, Lee et al.\cite{lee10} actually confirmed the theoretical prediction and
revealed ferroelectric and ferromagnetic order in the tensile strained EuTiO$_{3}$ thin
films deposited on DyScO$_{3}$ substrates. The possibility of inducing the ferroelectric
and ferromagnetic order in strained thin films of materials, which are paraelectric and
AFM in the bulk form, opens a new route for preparation of novel multiferroics with a
strong magnetoelectric coupling and with high critical ordering temperatures. Very
promising candidates are SrMnO$_{3}$,\cite{lee10b} EuO\cite{bousquet10} and
Ca$_{3}$Mn$_{2}$O$_{7}$.\cite{benedek11}

Until recently it was assumed that bulk EuTiO$_{3}$ has perovskite structure with cubic
$Pm\overline{3}m$ space group\cite{brous53} and that the structure is stable down to
liquid He temperatures. However, Rushchanskii et al.\cite{rushchanskii10,rushchanskii12}
theoretically investigated the structural and lattice dynamical properties of EuTiO$_{3}$
using first principles and revealed unstable phonons at the R
($\frac{1}{2},\frac{1}{2},\frac{1}{2}$) and M ($\frac{1}{2},\frac{1}{2},0$) points of the
Brillouin zone (BZ) in the cubic structure. Calculated eigenvectors indicated that the
instabilities are non-polar and arise from the tilting and rotation of the oxygen
octahedra. The eigenvectors for the M-point instable phonon show in-phase rotations of
the oxygen octahedra around one or more pseudocubic axes, whereas at the R point the
octahedra rotate with an alternating out-of-phase sense.\cite{rushchanskii12} Total
energy of the possible distorted phases were calculated and three possible stable
structures with $R\bar{3}c$, $Imma$ and $I4/mcm$ space group were suggested. The most
stable structure should be $I4/mcm$ ($a^{0},a^{0},c^{-}$ tilts in Glazer notation),
second stable structure could be $Imma$ ($a^{0}b^{-}b^{-}$) and the third one the
$R\bar{3}c$ structure ($a^{-}a^{-}a^{-}$). However, the energy differences between all
the structures are very small (within $\sim$ 2 meV per formula units), which were within
the range of numerical errors. Therefore all the above mentioned structures could be
realistic and the structural verification is needed. Very recent specific heat anomaly
measurements revealed an anomaly near 280\K, but the symmetry of the low-temperature
structure was not determined.\cite{Bussmann-Holder11} Allieta \textit{et
al.}\cite{allieta12} found the structural phase transition at 235\K\ and determined the
low-temperature structure as tetragonal $I4/mcm$. In this paper we will show that
critical temperature (T$_{c}$) of the antiferrodistortive phase transition strongly
depends on a quality of the EuTiO$_{3}$ samples. X-ray diffraction (XRD) and dynamical
mechanical analysis of the best EuTiO$_{3}$ ceramics reveals T$_{c}$=308\K, but XRD of
single crystal does not resolve the tetragonal symmetry down to 100\K. On other hand the
electron diffraction reveals tetragonal structure and moreover an incommensurate
modulation in single crystal already at room temperature. The reasons for such peculiar
effects will be discussed in details.

\section{Experimental}

We have investigated single crystals and two kinds of ceramics obtained by different
methods. At the beginning the EuTiO$_{3}$ powder was prepared from Eu$_{2}$O$_{3}$ and
Ti$_{2}$O$_{3}$ powders. The initial powder was pelletized and sintered at 1400 $^{o}$C
for 2 hours in a pure hydrogen atmosphere. Relative density of such prepared A ceramics
was 89\% of the theoretical one. Details of the A ceramic preparation are described
elsewhere.\cite{kachlik12} Ceramics B were prepared from exactly the same EuTiO$_{3}$
powder as the ceramics A, but the powder was loaded into Au capsules and sintered in a
belt-type high-temperature high-pressure furnace at 900 $^{\circ}$C under a pressure of 6
GPa for 30 min. Density of the ceramics B was more than 95\%.

Single crystals were prepared in two steps. Firstly, a mixture of stoichiometric amounts
of Eu$_{2}$O$_{3}$ (99.9 \% purity; \textit{Metall Rare Earth Limited}) and TiO$_{2}$ (99
- 100.5 \%; \textit{Sigma-Aldrich}) was ball-milled and sintered for 10 hours at 1273\K\,
under reducing atmosphere (flowing mixture of 5 \% H$_{2}$ in Ar$_{2}$, 100 ml/min). The
resulting phase was cubic-perovskite with a = 3.905 \AA. Secondly, the milled powder was
pressed into rods with 7 mm diameter and annealed for further 10 hours under the same
atmosphere. The crystals were grown under flowing mixture of 5 \% H$_{2}$ in Ar (150
ml/min) by using a floating-zone furnace equipped with four halogen lamps (maximum power
of 1500 W) and ellipsoidal mirrors. The obtained black crystals after polishing were
porosity free with mirror quality surface. However, the crystals were not perfect, they exhibited mosaicity
in the mm range size. One crystal was grinded for powder X-ray and electron
diffraction studies after dielectric, magnetic and infrared (IR) measurements. For IR
studies the crystal with the size of 3x3x0.15 mm$^{3}$ was generally oriented with [001]
axis tilted approximately 30$^{\circ}$ from the sample normal plane. Therefore the IR
spectra were taken without a polarizer (polarized IR spectra did not show any
anisotropy).

The X-ray diffraction studies of ceramics were performed using a Bruker D8 Discover
equipped with rotating Cu anode ($\lambda$(Cu$K\alpha_1$)=1.540598 \AA;
$\lambda$(Cu$K\alpha_2$)=1.544426 \AA) working with 12 kW power. Parabolic G\"{o}bel
mirror was located on the side of the incident beam. Analyzer slits and alternatively
also analyzer crystal (200 - LiF) were on the side of the diffracted beam. The
temperature was controlled by cooling stage Anton Paar DCS 350. The temperature was
changed from 173\K\, to 373\K. The X-ray $2\Theta$/$\Theta$ diffractograms were measured
in the broad range of 2$\Theta$ angles from 25 till 135 $^{\circ}$ at 193, 293 and 333\K.
These whole scans were used for space group determination and for Rietveld refinement
with program Topas.\cite{topas} Although the analyzer crystal reduces the intensity, it
significantly improves the resolution. Thus, it was used for detailed study of the 310
and 420 diffraction peaks (using cubic indexes) that were measured almost each 10\K\, between
173 and 373\K. Below 300\K, the Rietveld refinements of these two diffraction peaks were
performed in $I4/mcm$ space group, because the R$_{wp}$ factor was noticeably lower than
in $Pm\bar{3}m$ space group. The fixed microstructure parameters provide precise values
of lattice parameters that were used for calculation of the oxygen octahedra tilting
angle $\phi$ :\cite{mitchell02}
\begin{equation}\label{uhel-tiltu}
cos\phi=\frac{\sqrt{2}a_{tetr}}{c_{tetr}}
\end{equation}

X-ray diffraction studies of EuTiO$_{3}$ single crystal were performed with four-circle
kappa diffractometer Gemini of Oxford Diffraction (now Agilent Technologies), equipped
with CCD detector Atlas. Because of large absorption of the sample we used MoK$\alpha$
radiation with doublet, $\lambda$=0.7107 \AA, monochromatized with a graphite
monochromator and collimated with a fibre-optics Mo-Enhance collimator of Oxford
Diffraction. The temperature was controlled with an open-flow cooler Cryojet HT of Oxford
Instruments, which uses nitrogen gas as a cooling/heating medium. The measurements were
performed at 300 and 100\K.

Electron diffraction patterns were collected at room temperature using transmission
electron microscope Philips CM120 equipped with CCD camera Olympus Veleta with 14 bit
dynamical range. Crystals of around 500 nm size were investigated. A tilt series of
diffraction patterns was recorded ranging from -50$^{\circ}$ to +50$^{\circ}$ in steps of
0.5$^{\circ}$. This technique is known as electron diffraction
tomography.\cite{gorelik11} Its main advantages are that it is very fast, does not
require the cumbersome procedure of orienting the crystal along some special zone axis,
and grant access to a full 3D intensity distribution in reciprocal space. Oriented
reciprocal-space sections were then reconstructed from the raw data by the computer
program PETS. \cite{palatinus}

The Young's modulus and thermal expansion were measured using the instrument for
Dynamical Mechanical Analysis (DMA 7 - Perkin Elmer). The parallel-plate-method,
described in details elsewhere,\cite{kityk96} was used. The sample dimensions of the parallelepiped
were 2.961x0.922x0.745 mm$^{3}$. A static force of 750 mN was modulated by a dynamic force of 700\,mN
with a frequency of 1 Hz. Nitrogen was used as purge gas.

Low-frequency (1 kHz - 1 MHz) dielectric measurements were performed between 2 and
300\K\, using NOVOCONTROL Alpha-A High Performance Frequency Analyzer. Magnetic
susceptibility data were obtained using a Quantum Design PPMS9.

The IR reflectivity spectra were taken using a Bruker IFS 113v FTIR spectrometer at
temperatures from 10 to 300\K\, with the resolution of 2\cm. An Optistat CF cryostat
(Oxford Instruments) was used for cooling the samples. The investigated spectral range
(up to 650\cm) was limited by the transparency region of the polyethylene windows of the
cryostat. A helium-cooled Si bolometer operating at 1.6\K\, was used as a detector.
Room-temperature reflectivity was measured up to 3000\cm\, using pyroelectric deuterated
triglicine sulfate detector.

IR reflectivity spectra were carefully fitted assuming the
dielectric function in the factorized form of generalized damped
harmonic oscillators\cite{gervais83}
\begin{equation}\label{eps4p}
\varepsilon^{*}(\omega)=\varepsilon_{\infty}\prod_{j=1}^n\frac{\omega^{2}_{LOj}-\omega^{2}+i\omega\gamma_{LOj}}{\omega^{2}_{TOj}-\omega^{2}+i\omega\gamma_{TOj}}
\end{equation}
where $\omega_{TOj}$ and $\omega_{LOj}$ denote the transverse and longitudinal frequency
of the j-th polar phonon, respectively, and $\gamma$$_{TOj}$ and $\gamma$$_{LOj}$ denote
their corresponding damping constants. $\varepsilon$$^{*}$($\omega$) is related to the
reflectivity R($\omega$) of the bulk substrate by
\begin{equation}\label{refl}
R(\omega)=\left|\frac{\sqrt{\varepsilon^{*}(\omega)}-1}{\sqrt{\varepsilon^{*}(\omega)}+1}\right|^2
.
\end{equation}
The high-frequency permittivity $\varepsilon_{\infty}$ = 5.88 resulting from the
electronic absorption processes was obtained from the room-temperature
frequency-independent reflectivity tails above the phonon frequencies and was assumed to
be temperature independent.

\begin{figure}
  \begin{center}
    \includegraphics[width=80mm]{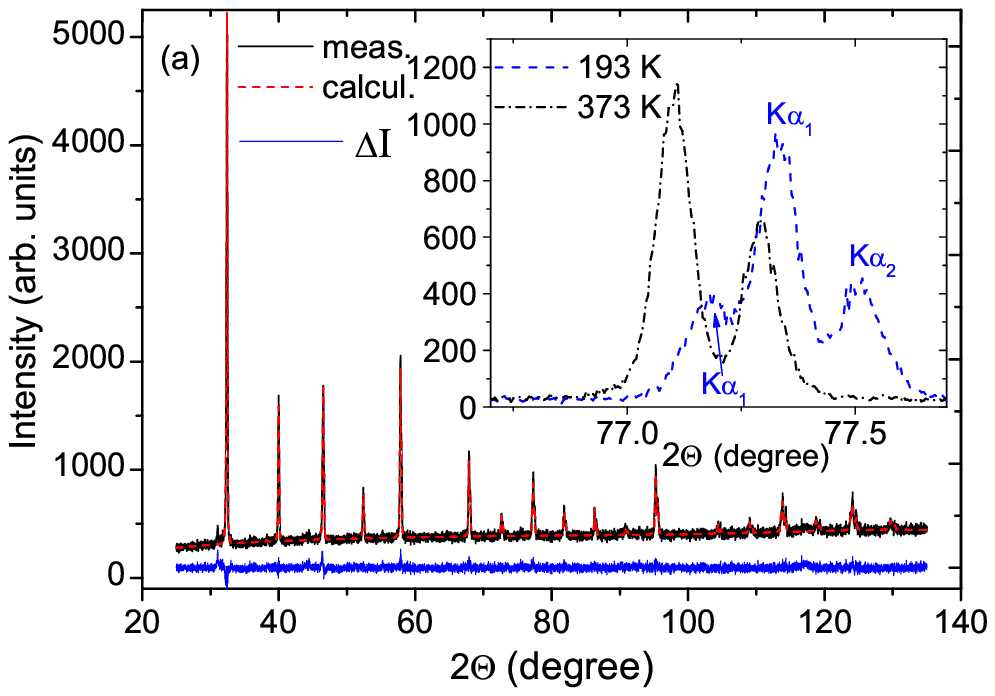}
    \includegraphics[width=75mm]{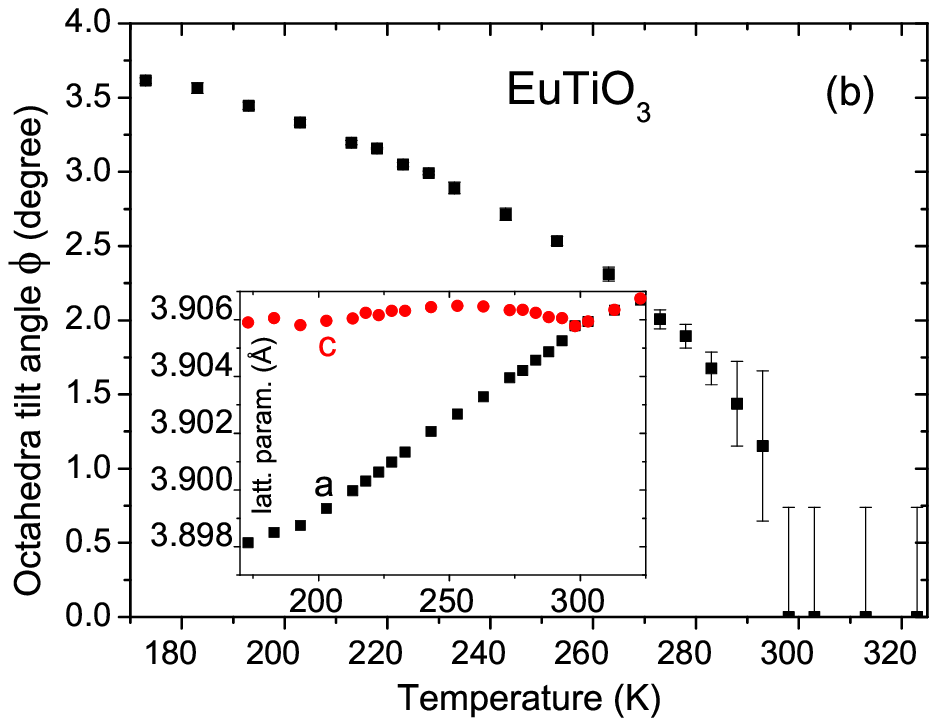}
  \end{center}
    \caption{(Color online) a) Measured, calculated and residual x-ray diffraction patterns at 300\K\ for sample A.
    The inset shows the splitting of 310 reflection peak on cooling from 373 down to 193\K. Note the
    $\alpha_{1}$+$\alpha_{2}$ doublet of 310 reflection. b) Temperature dependence of tilt angle of oxygen octahedra
    from the \textbf{c} axis.
    Below 260\K\, the error bars are smaller than the points .
    Inset shows temperature dependence of pseudocubic lattice parameters $a$ and $c$. Real tetragonal parameters
    are defined as follows: $a_{tetr}=\sqrt{2}a$, $c_{tetr}=2c$.}
    \label{Fig1}
\end{figure}

\section{Results}

\subsection{Structural and elastic properties of the A ceramics}

Above room temperature the X-ray diffraction (see Fig.~\ref{Fig1}a) confirms cubic
$Pm\bar{3}m$ structure, which is in complete agreement with Ref. \cite{brous53}. Below
300\K\ the structure was refined in $I4/mcm$ space group, because R$_{wp}$ factor was
noticeably reduced in this structure. For example at 273\K\, R$_{wp}$ was 9.12 and 12.90
in tetragonal and cubic structure, respectively. Tetragonal $I4/mcm$ structure is
obtained by an anti-phase tilting of oxygen octahedra along the c axis ($a^{0}a^{0}c^{-}$
in Glazer\cite{glazer74} notation). The structure models in $R\bar{3}c$ and $I mma$ space
groups were rejected, because of discernible discrepancies in the Rietveld refinements.
Due to the tetragonal distortion the lattice parameter splits (see inset of
Fig.~\ref{Fig1}b). The low-temperature value of the splitting is comparable to the
theoretical value in Ref. \cite{rushchanskii12}. Allieta \textit{et al.} observed the
phase transition only at 235\K, but they obtained comparable lattice parameter splitting
$\sim$ 130\K\, below T$_{c}$ as we, just their lattice parameters were systematically
$\sim$ 0.003 $\AA$ smaller than ours. The temperature dependence of the part of the
diffraction pattern close to the 310 reflection is shown in the inset of
Fig.~\ref{Fig1}a. It is a doublet at high temperatures due to the K$\alpha1\alpha2$ lines
of Cu, which splits on cooling due to the antiferrodistortive transition. It allowed us
to determine the temperature dependence of the tilting angle from the c axis
(Fig.~\ref{Fig1}b). One can see that the antiferrodistortive phase transition arises
between 295 and 320\K\, and the tilt angle $\phi$ reaches a value of 3.6$^{\circ}$ at
173\K. The precision of lattice parameters determination was around 0.0001 \AA. For very
close $a$ and $c$ lattice parameters it leads to relatively big errors in the
determination of $\phi$, see equation (1). For these reasons the error bars near and
above 300\K\, in Fig.~\ref{Fig1}b are large. At low temperatures the accuracy of the tilt
angle determination is much higher and the error bars are smaller than the dots.

\begin{figure}
  \begin{center}
    \includegraphics[width=50mm]{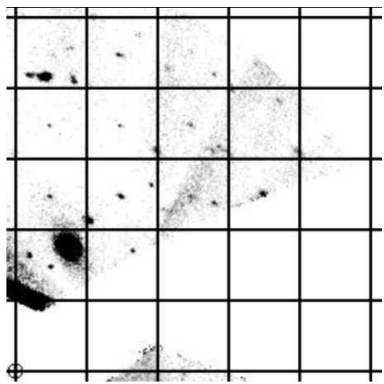}
  \end{center}
    \caption{A reconstructed section of the diffraction pattern of ceramics A showing the layer $\frac{1}{2}$kl.
    The pseudocubic reciprocal lattice is outlined with grid, point $\frac{1}{2}$00 is marked with a circle
    in the lower left corner. Additional spots not located at the positions
    $\frac{1}{2}$,k+$\frac{1}{2}$,l+$\frac{1}{2}$ come from another domain in the crystal.}
    \label{Fig2}
\end{figure}

We checked the structure also by electron diffraction on small single-crystalline grains
obtained by grinding of A ceramics. The electron diffraction pattern taken at 295\K\,
contains weak spots at the position h+$\frac{1}{2}$,k+$\frac{1}{2}$,l+$\frac{1}{2}$ (see
Fig.~\ref{Fig2}). It confirms tetragonal distortion in ceramics A at room temperature.

\begin{figure}
  \begin{center}
    \includegraphics[width=80mm]{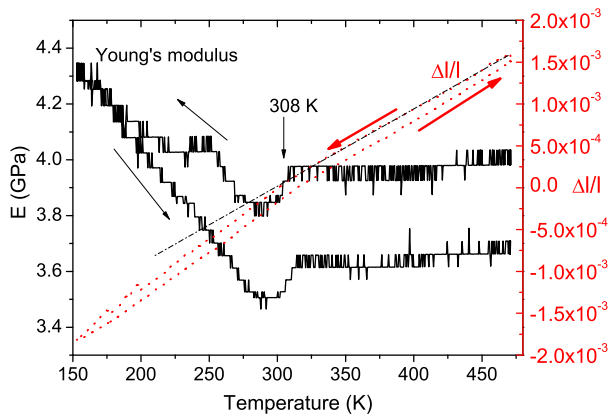}
  \end{center}
    \caption{(Color online) a) Temperature dependence of Young's modulus $E$ and thermal expansion
    $\Delta l/l$ in bulk EuTiO$_{3}$ ceramics sintered by the A method. }
    \label{Fig3}
\end{figure}

The existence of a structural phase transition was confirmed also by other methods. In
Fig.~\ref{Fig3} we show the temperature dependencies of the thermal expansion and Young's
modulus. The curves were systematically measured first on cooling and then on heating. As
Fig.~\ref{Fig3} shows the thermal expansion depends linearly on temperature above and
below 308\K, indicating that the phase transition at 308 K is of second order. Similar as
\textit{e.g.} for SrTiO$_{3}$ \cite{kityk00} or KMnF$_{3}$ \cite{schranz09} the thermal
expansion in EuTiO$_{3}$ is caused by the coupling term which is quadratic in the order
parameter and linear in the strain $\epsilon$ in a Landau Free energy expansion, implying
for the spontaneous strain $\epsilon_{s}$ to be proportional to the square of the order
parameter $\eta$, i.e. $\epsilon_{s} \sim \eta^{2} \sim (T_{c}-T)$ for a second order
phase transition. This is also consistent with the observed anomaly in the Young´s
modulus, which displays a negative dip at $T_{c}$ followed by a linear increase with
decreasing temperature. To describe the elastic anomaly in EuTiO$_{3}$ we can similarly
to SrTiO$_{3}$ \cite{kityk00} employ the leading coupling terms $\sim a \eta^{2}\epsilon
+ b \eta^{2}\epsilon^{2}$, which leads to
\begin{equation}\label{uhel-tiltu}
E=E^{0}-\frac{2a^{2}}{B}+b\eta^{2} \propto E^{0}-\frac{2a^{2}}{B}+(T_{c}-T)
\end{equation}
in perfect agreement with observed behavior (Fig.~\ref{Fig3}).

Similar DMA experiments were performed on several EuTiO$_{3}$ ceramics as well as on
ceramics with some pyrochlore or amorphous impurities (all samples prepared by the A
method). The phase transition was always observed, but its temperature was reduced in
dependence of the impurity concentration. Maximal shift down of the critical temperature
was 60\K\, in samples with $\sim$ 2\% of pyrochlore impurities. Such a low phase
transition temperature has been very recently reported by Allieta et al.\cite{allieta12}.
In the high resolution synchrotron X-ray powder diffraction they found a signature of the
phase transition in EuTiO$_{3}$ at 235\K. Influence of defects on structural properties
of EuTiO$_{3}$ will be discussed in details in the next paragraph.

\subsection{X-ray diffraction studies of the B ceramics}

\begin{figure}
  \begin{center}
    \includegraphics[width=70mm]{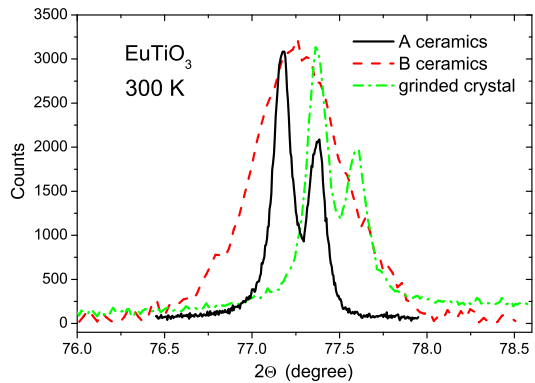}
  \end{center}
    \caption{Room-temperature 310 powder diffraction peak of the A and B ceramics compared with the same
    diffraction peak of grinded single crystal. Higher angle of the diffraction in single crystal gives evidence
    about its smaller lattice constant than in ceramics. The width of diffraction is broader in the B ceramics
    due to its smaller crystallite size and larger microstrain. Intensities of all diffractions are normalized
    for their easier comparison.}
    \label{Fig4}
\end{figure}

The B ceramics was sintered at 900\,$^{\circ}$C (and at pressure 6 GPa) while the A
ceramics was sintered at 1400\,$^{\circ}$C (and at ambient pressure). For that reason
smaller crystallite size and larger microstrain (i.e. fluctuation of lattice parameters)
is expected in the B ceramics. In Fig.\ref{Fig4} is compared 310 diffraction peak of A and B
ceramics together with grinded single crystal. One can see that the diffraction of the B
ceramics is very broad and therefore does not allow to resolve the K$\alpha_1$ and
K$\alpha_2$ splitting. From the width of the diffraction we determined the crystallite
size 40\,nm and microstrain 0.2\%, while in the A ceramics we obtained the crystallite
size 90 nm and microstrain 0.03\%. Due to large width of diffraction peaks no structural
phase transition could be resolved in the B ceramics down to 100\K. However, we will show
below, that IR spectra give evidence about lower than cubic structure in the B ceramics
at all temperatures below 300\K.

\subsection{X-ray and electron diffraction studies of EuTiO$_{3}$ single crystal}

Two small single crystals broken from one larger piece were used for the X-ray
diffraction studies. Right-angle triangle with sides 0.2\,mm and thickness 0.08\,mm
showed two grains, while the triangular-shaped sample with sides 0.7\,mm and thickness
0.2\,mm contained 9 single-crystal grains. Within its accuracy and sensitivity, the
laboratory single-crystal X-ray diffraction experiment did not reveal any tetragonal
distortion or presence of additional spots, as observed in the A ceramics. The cell
parameters determined at 100\K\, were a=3.89114(11) \AA, b=3.89144(11) \AA,
c=3.89117(10)\AA, $\alpha$=89.984(2)$^{\circ}$, $\beta$=90.017(2)$^{\circ}$,
$\gamma$=89.984(2)$^{\circ}$ The structure was therefore refined in a cubic space group,
yielding R value 0.85\% for 74 observed symmetry independent reflection, with goodness of
fit 1.07.
 Important for successful structure refinement was correction for exceptionally strong extinction.
Many spots, which could be explained by a doubling of the unit cell in the
antiferrodistortive phase, were observed, but finally explained by $\lambda$/2
diffraction. After decreasing the voltage on the Mo X-ray tube below the $\lambda$/2
generation limit, \textit{i.e.} below 34 kV, no such satellite appeared although the loss
of intensity due to the decrease of the voltage was fully compensated by increasing mA
rate and the exposition time.

\begin{figure}
  \begin{center}
    \includegraphics[width=50mm]{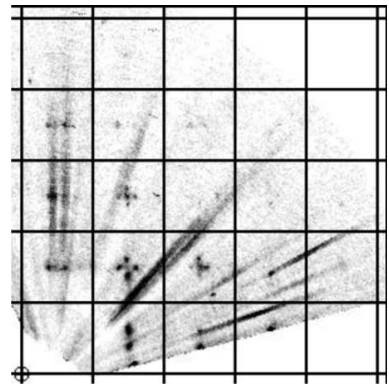}
  \end{center}
    \caption{A reconstructed section of the diffraction pattern of a sample prepared from a monocrystal showing the
    layer $\frac{1}{2}$kl. The pseudocubic reciprocal lattice is outlined with grid, point $\frac{1}{2}$00 is marked
    with a circle in the lower left corner. Streaks in the image are traces of thermal diffuse scattering coming from
    reflections outside the depicted plane.}
    \label{Fig5}
\end{figure}

The second larger crystal was grinded two hours after the X-ray measurements at 100\K\,
and a selected grain of the powder was used for electron diffraction at room temperature.
The diffraction pattern (see Fig.~\ref{Fig5}) reveals weak reflections at
positions h+$\frac{1}{2}$,k+$\frac{1}{2}$,l+$\frac{1}{2}$, which are evidence for
tetragonally distorted phase like in the ceramics A. Moreover, the satellites around these
positions are clearly seen in Fig.~\ref{Fig5}. The satellites show that the structure is
incommensurately modulated with modulation wavevector \textbf{q$_{m}$} = (0.38 $\pm$ 0.02)
\textbf{a}$^*$.

\begin{figure}
  \begin{center}
    \includegraphics[width=70mm]{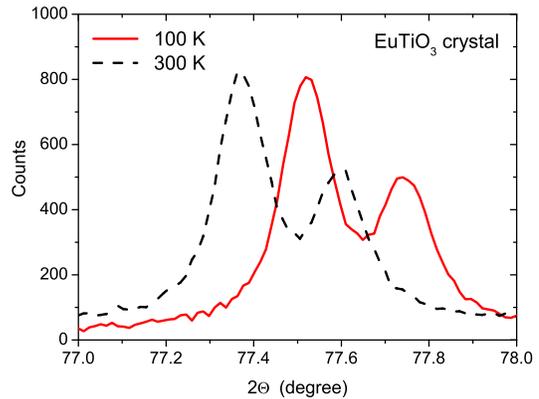}
  \end{center}
    \caption{The 310 diffraction of grinded single crystal at room temperature and at 100\K.
    No line broadening or splitting is seen on cooling,
    i.e. no evidence of tetragonal distortion is revealed. }
    \label{Fig6}
\end{figure}

We repeated the electron diffraction measurements on the same sample two weeks later. No
reflections at positions h+$\frac{1}{2}$,k+$\frac{1}{2}$,l+$\frac{1}{2}$ and as well as
no incommensurate satellites were observed. The crystal lattice looked cubic. Very recent
synchrotron study of Kim \textit{et al.}\cite{kim12} revealed creation of incommensurate
structure in EuTiO$_{3}$ crystal at 285\K\, and the tetragonal distortion appeared only
below 160\K. The incommensurate phase transition is of the first order, therefore it can
exhibit some temperature hysteresis. Our first electron diffraction pattern was taken
only two hours after cooling to 100\K, therefore we observed the tetragonal distortion
and incommensurate modulation at room temperature. After some time the crystal structure
transforms to cubic one at 300\K. We observed temperature hysteresis in tilt angle of
oxygen octahedra as well as in T$_{c}$ with value of 30-50\K\, also in the A ceramics.

Our modulation wavevector \textbf{q$_{m}$} = 0.38 \textbf{a}$^*$ is smaller than the
value 0.43 \textbf{a}$^*$ reported by Kim \textit{et al.}\cite{kim12}, but Kim \textit{et
al.} have also shown that the incommensurate satellites are strongly time dependent
(their position and intensity strongly relaxed within measured 17 hours). The time and
possible temperature dependence of the modulation wavevector is probably responsible for
the discrepancy between ours and Kim's value of \textbf{q$_{m}$}.

Nevertheless, the discrepancy between data obtained using XRD of macroscopic crystal and
electron diffraction of the same grinded crystal are remarkable and one could speculate
about lower sensitivity of the former method. Therefore we have decided to use the same
grinded crystal for powder X-ray diffraction measurements. The results were rather
surprising (see Figs. \ref{Fig4} and \ref{Fig6}): a) The room-temperature lattice
constant of the crystal (a=3.8966(5) $\textrm{\AA}$) is lower than in the A ceramics
(a=3.9058(5) $\textrm{\AA}$). b) Diffraction peaks (like 310 in Fig.~\ref{Fig6}) do not
broaden or split on cooling down to 100\K. It means the lattice of the crystal looks
cubic down to 100\K. Note the electron diffraction revealed the incommensurate satellites
and tetragonal distortion in grinded single crystal already at room temperature (after
cooling to 100\K). Electrons interact much stronger with the crystal than X-rays, and the
weak intensities can be also enhanced by the dynamical diffraction effects. This could
explain the discrepancy between both kinds of experiments. Here should be again stressed
that the recent X-ray synchrotron investigation of EuTiO$_{3}$ single crystal, which has
the same origin as ours, revealed the tetragonal distortion below 160\K\, and
incommensurate modulation below 285\K.\cite{kim12}

One can also ask the question, whether the discrepancies between XRD and electron
diffraction of the crystals cannot be caused by stresses created after grinding of the
crystals. It is unlikely that such manipulation would alter the crystal structure. Also,
if the material was affected by the crushing, the diffraction pattern would probably show
streaking and loss of crystallinity. Moreover, the structure would be most probably
different from one grain to another, but we observed the same electron diffraction
patterns in three examined crystals. We proved that the material relaxed to its cubic
form without satellites in two weeks after the first experiment, indicating strongly a
temperature hysteresis.

We tried to see the phase transitions in specific heat of the samples. Unfortunately, we
did not detect any anomaly. Our differential scanning calorimeter Perkin Elmer Pyris
Diamond has probably lower sensitivity than the instrument used in Ref.
\cite{Bussmann-Holder11}

\begin{figure}
  \begin{center}
    \includegraphics[width=75mm]{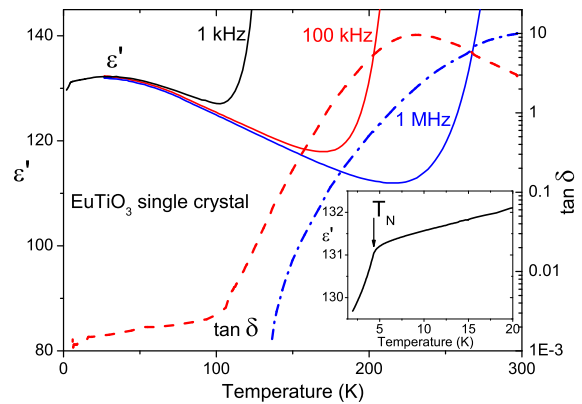}
  \end{center}
    \caption{Temperature dependence of dielectric permittivity and loss in EuTiO$_{3}$ single crystal taken
     at various frequencies. Inset shows decrease of 1 kHz permittivity below N\'{e}el temperature due to
     spin-phonon coupling. }
    \label{Fig7}
\end{figure}

\begin{figure}
  \begin{center}
    \includegraphics[width=60mm]{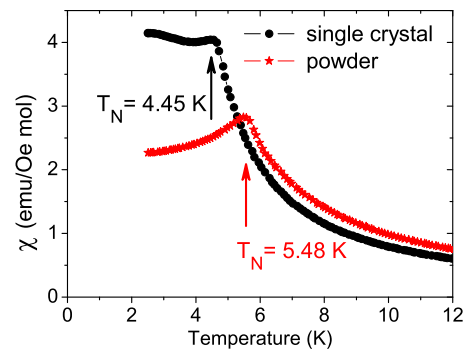}
  \end{center}
    \caption{Temperature dependence of magnetic susceptibility in EuTiO$_{3}$ powder and single crystal.}
    \label{Fig8}
\end{figure}

Not only structural but also dielectric and magnetic properties differentiate in various
samples. A ceramics has at low temperatures permittivity about 400 (like single crystal
in Ref. \cite{katsufuji01}), B ceramics about 200 and the single crystal less than 140
(see Fig.~\ref{Fig7}). The resistivity was highest in the single crystal, which allowed
to measure intrinsic dielectric permittivity at 1\,MHz up to 200\K\, (Fig.~\ref{Fig7}),
while in the A ceramics permittivity was affected by Maxwell-Wagner polarization already
above 80\K.\cite{kamba07} We measured as well magnetic susceptibility and found T$_{N}$ =
4.45\K\, in single crystal, and T$_{N}$=5.48\K\, in the EuTiO$_{3}$ ceramics and powder,
from which the crystal was grown (see Fig.~\ref{Fig8}). The latter value corresponds to
the value reported in the literature. \cite{katsufuji01,guire66,kamba07} The reason for
lower T$_{N}$ in single crystal is discussed below.

Our A ceramics exhibits the antiferrodistortive phase transition to tetragonal $I4/mcm$
structure at $\sim$ 300\K, K\"{o}hler \textit{et al.} reported T$_{c}$ =
282\K,\cite{Bussmann-Holder11,kohler12}, Allieta \textit{et al.} found T$_{c}$
=235\K,\cite{allieta12} and Kim \textit{et al.} at 160\K. All the discrepancies in
critical temperatures (as well as in above mentioned dielectric and magnetic properties)
observed in different samples can be explained only by a different concentration of
defects (mainly oxygen vacancies and Eu$^{3+}$). It is known that the oxygen vacancies
enhance the lattice constant in isostructural SrTiO$_3$. \cite{gong91} On other hand,
ionic radius of Eu$^{3+}$ is smaller than ionic radius of Eu$^{2+}$,\cite{jia91} so the
Eu$^{3+}$ defects can reduce the lattice constant in Eu$^{2+}$TiO$_{3}$. One can expect
that both kinds of defects are mutually connected. Oxygen vacancies can strongly
influence the lattice instability to tetragonal phase, where the oxygen octahedral
exhibit antiphase tilting. Note as well that Allieta et al.\cite{allieta12} observed
local fluctuations of the tilt angle, which can be explained by fluctuation of oxygen
vacancy concentration. On other hand the Eu$^{3+}$ can also reduce the N\'{e}el
temperature in EuTiO$_{3}$. It can explain aforementioned 1\K\, lower T$_{N}$ in single
crystal than in the ceramics. For that reason it seems that the EuTiO$_{3}$ crystal
contains more Eu$^{3+}$ defects than the A and B ceramics. Exact determination of
concentration of Eu$^{3+}$ and oxygen vacancies using M\"{o}ssbauer spectroscopy and
positron annihilation spectroscopy, respectively, is beyond the scope of this paper, but
these experiments are already in progress in our lab.

\subsection{Factor-group analysis}

As it was already mentioned above, Rushchanskii et al.\cite{rushchanskii12} predicted
three possible space groups, in which EuTiO$_{3}$ could crystallize at low temperatures.
Structural analysis can yield sometimes ambiguous results and IR spectra can help in
specification of the crystal structure. Before presenting the IR spectra, we perform
the factor group analysis of the optical phonons (\textit{i.e.} without acoustic modes) in all
suggested crystal structures of EuTiO$_{3}$. In cubic $Pm\overline{3}m$ structure the
analysis gives the following symmetries of phonons in the $\Gamma$-point of BZ:
\begin{equation}\label{cubic-analysis}
\Gamma_{Pm\overline{3}m}=3F_{1u}(x)+F_{2u}(-).
\end{equation}
$x$ in bracket means activity in the IR spectra, (-) marks the silent mode. The
factor-group analysis in antiferrodistortive tetragonal $I4/mcm$ phase gives the
following:
\begin{eqnarray}\label{tetragonal-analysis}
\Gamma_{I4/mcm}&=&5E_{u}(x,y)+3A_{2u}(z)+A_{1g}(x^{2}+y^{2},z^{2})+ \nonumber\\
& & 3E_{g}(xz,yz)+2B_{1g}(x^{2}-y^{2})+2A_{2g}(-)+\nonumber\\
& & B_{2g}(xy)+A_{1u}(-)+ B_{2u}(-).
\end{eqnarray}
Here $x^{2},xy$ etc. mean the components of Raman tensor, where the modes are Raman
active. It follows from the analysis that eight polar phonons are expected in IR spectra
of tetragonal EuTiO$_{3}$. Their theoretical frequencies and oscillator strengths are
listed in Table \ref{phonon-freq}. In the orthorhombic $Imma$ phase the factor group
analysis yields
\begin{eqnarray}\label{orthorhom-analysis}
\Gamma_{Imma}& = & 5B_{1u}(z)+4B_{2u}(y)+4B_{3u}(x)+3A_{g}(x^{2})+ \nonumber\\
& & 3B_{2g}(xz)+4B_{3g}(yz)+2A_{u}(-)+2B_{1g}(xy),\nonumber\\
& &
\end{eqnarray}
while in rhombohedral phase the optical phonons have the following symmetry in the center
of BZ:
\begin{eqnarray}\label{rhombo-analysis}
\Gamma_{R\bar{3}c}& = & 5E_{u}(x,y)+3A_{2u}(z)+A_{1g}(x^{2}+y^{2},z^{2})+\nonumber\\
& & 4E_{g}(xz,yz,x^{2}-y^{2},xy)+3A_{2g}(-)+2A_{1u}(-).\nonumber\\
& &
\end{eqnarray}
It means that 13 and 8 IR active phonons are expected in the
orthorhombic and rhombohedral phase of EuTiO$_{3}$, respectively.
Let us compare the experimentally observed phonons in various
EuTiO$_{3}$ ceramics and single crystals with the predicted
selection rules mentioned above.

\begin{figure}
  \begin{center}
    \includegraphics[width=70mm]{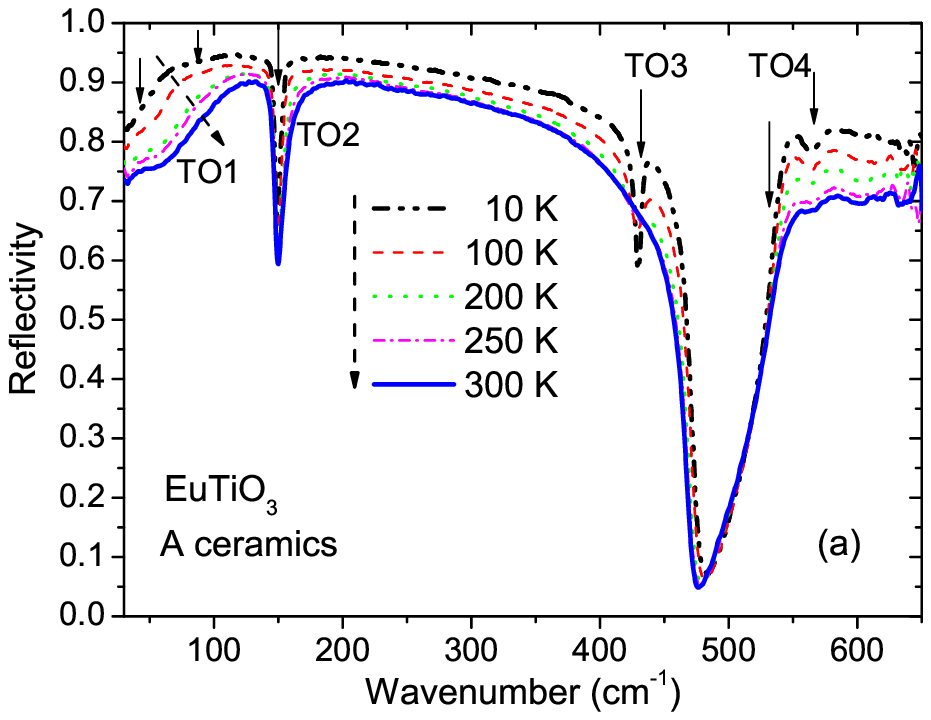}
    \includegraphics[width=70mm]{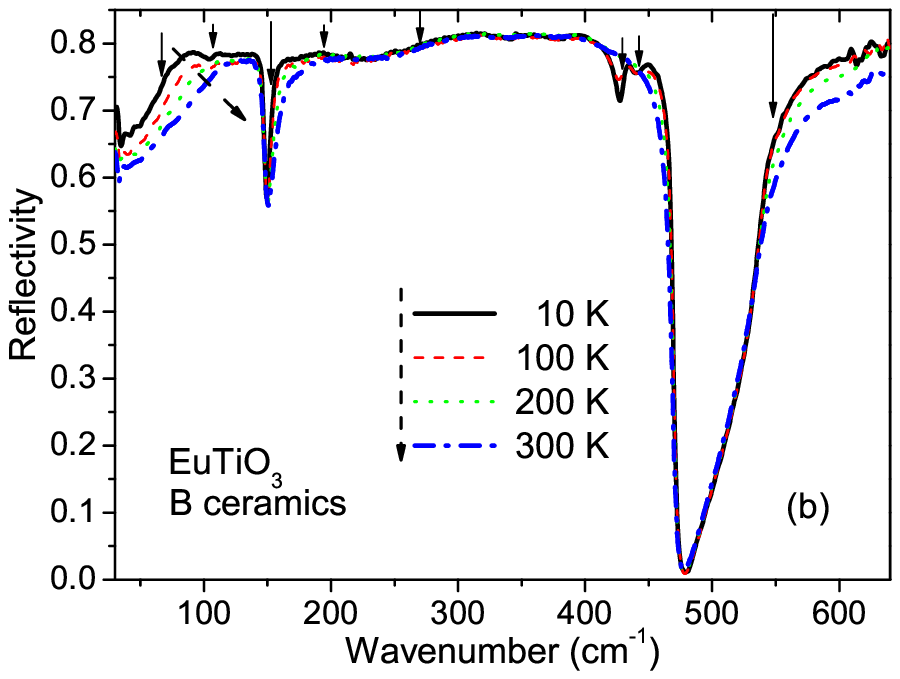}
    \includegraphics[width=70mm]{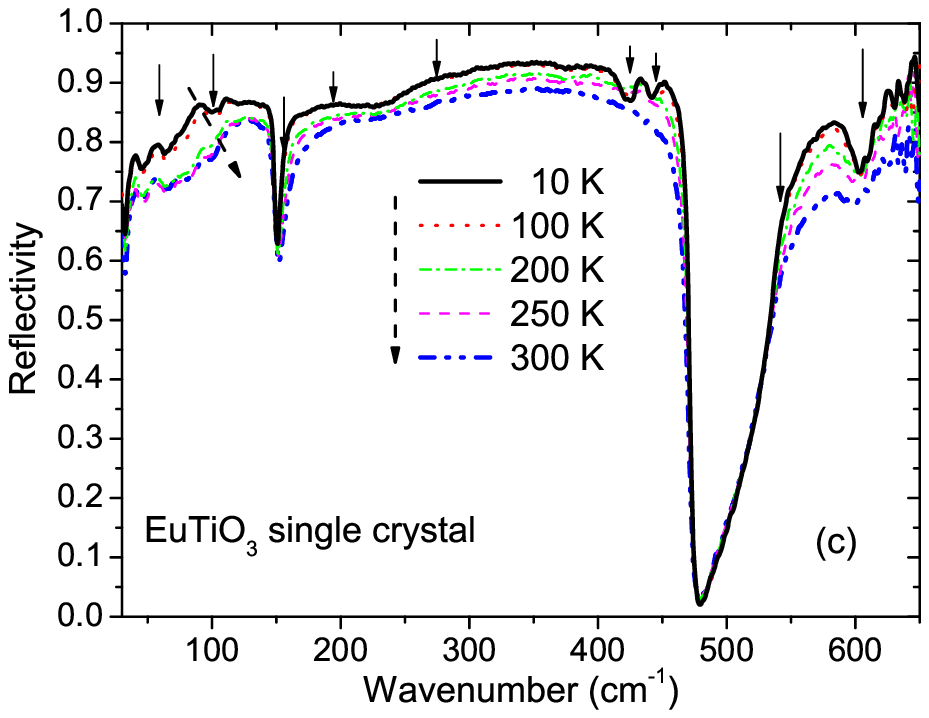}
  \end{center}
    \caption{(Color online) a) Temperature dependence of IR reflectivity in a) A ceramics, b) B ceramics and
    c) single crystal of
     EuTiO$_{3}$. Solid arrows mark frequencies of polar phonons. Artificial low-frequency oscillations
     in the single-crystal spectra are caused by a small size of the crystal. }
    \label{Fig9}
\end{figure}

\subsection{IR studies}

\begin{figure}
  \begin{center}
    \includegraphics[width=70mm]{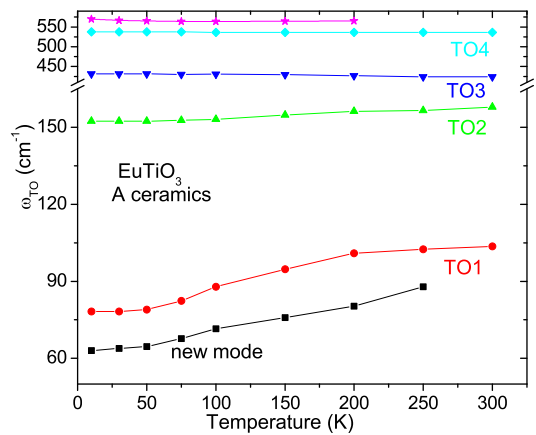}
  \end{center}
    \caption{(Color online) Temperature dependence of the polar phonon frequencies in the A
    ceramics. }
    \label{Fig10}
\end{figure}

Fig.~\ref{Fig9} compares the IR reflectivity spectra of the A and B EuTiO$_{3}$ ceramics
with the spectra of a single crystal. Note the similarity between the B ceramics and the
single crystal, while the A ceramics has different spectra. Shape of IR spectra of the A
ceramics is similar to the previously published spectra.\cite{kamba07,goian09} At room
temperature three reflection bands marked as TO1, TO2 and TO4 correspond to 3$F_{1u}$
symmetry modes of the cubic $Pm\overline{3}m$ structure (see Eq.~\ref{cubic-analysis}).
However, at low temperatures TO1 and TO4 modes split and moreover additional new mode
activates in the spectra near 430\cm\, (see Figs.~\ref{Fig9} and ~\ref{Fig10}). Hints of
these new modes we observed already in Ref. \cite{kamba07}, but that time we interpreted
them as impurity modes from the pyrochlore Eu$_{2}$Ti$_{2}$O$_{7}$ second phase. However,
our new spectra obtained from phase pure EuTiO$_{3}$ A ceramics show the new modes with
even higher intensities. Moreover, as we know that the structure of the A ceramics is
tetragonal below 300\K, the IR selection rules must be changed (see
Eq.~\ref{tetragonal-analysis}). All TO modes should be split and, moreover, two new
E$_{u}$ symmetry modes should be activated. The new sharp mode seen near 430\cm\, comes
from a silent TO3 mode (originally of $F_{2u}$ symmetry in the cubic phase) and it has
the $E_{u}$ symmetry in tetragonal phase (see Eq.~\ref{tetragonal-analysis}). Another
polar mode should be activated around 250\cm,\cite{rushchanskii12}, but its theoretical
strength is one order of magnitude lower than that of the other modes (see
Table~\ref{phonon-freq}). For this reason the mode is not resolved in our spectra.

Phonon eigenfrequencies $\omega_{TO}$ and mode plasma frequencies
$\Omega_{P}=\sqrt{\Delta\varepsilon}\omega_{TO}$ of the observed polar phonons in all
investigated samples are listed in Table~\ref{phonon-freq}. These experimental parameters
are compared with theoretical values obtained from first principles\cite{rushchanskii12}
in tetragonal, orthorhombic and cubic structure. If we neglect a small theoretical
splitting of TO2 mode, which lies below our spectroscopic resolution, very good agreement
of the experimental phonon parameters in the A ceramics and theoretical parameters in the
tetragonal $I4/mcm$ structure was obtained.

TO1 phonon splitting was resolved at 250\K\, \textit{i.e.} $\sim$ 50\K\, below T$_{C}$
(see Fig.~\ref{Fig10}), because the order parameter slowly increases on cooling below
T$_{C}$. The order parameter exhibits the same temperature dependence as the tilt angle
in Fig.~\ref{Fig1}b. The new mode has a lower frequency than the original one and softens
down to 60\cm\,(see Fig.~\ref{Fig10}). This new mode was not mentioned in previous
publications,\cite{kamba07,goian09} because just a small hint was seen in the IR
reflectivity band near 90\cm\, (Fig. ~\ref{Fig9}a). However, if the TO1 mode is fitted
with one oscillator only, its damping increases on cooling,\cite{goian09} which is not
physically meaningful. In our two-component fit of the TO1 band, both phonon dampings
decrease on cooling, which is reasonable.

TO3 mode is seen like a reflectivity minimum near 430\cm\, (see Fig. ~\ref{Fig9}a) mainly
at low temperature, but the reflectivity fits needs this mode at all temperatures up to
room temperature. Only the damping of this mode is significantly enhanced at high
temperatures, therefore no clear minimum (just sagging) is seen in reflectivity near
430\cm\, close to room temperature. It supports our results in paragraph A that the A
ceramics crystallizes in tetragonal structure already at 300\K.

The B ceramics and single crystal exhibit a very similar IR reflectivity spectra, but
different than the A ceramics (see Fig.~\ref{Fig9}). All together 9 modes were observed
in the single crystal spectra. In comparison to the A ceramics, two additional phonon
bands are seen in single crystal between 200 and 270\cm\, (manifested by bending of
reflectivity around 230\cm). Moreover, a doublet (instead of a singlet in the A ceramics)
arises below 450\cm\, at low temperatures. Their frequencies correspond well to IR active
phonons obtained theoretically in the orthorhombic $Imma$ structure \cite{rushchanskii12}
(see also Table~\ref{phonon-freq}). Kim \textit{et al.}\cite{kim12} claim, based on their
X-ray synchrotron data, that the structure locally fluctuates between antiferrodistortive
and antiferroelectric structure, so one can expect lower than tetragonal symmetry. Our
X-ray and electron diffractions do not support the orthorhombic structure, but one should
admit, that our electron diffraction measurement was not performed below room
temperature. Nevertheless, the additional polar phonons seen in the B ceramics and single
crystal can be activated due to incommensurately modulated structure. In such structure
the translation symmetry is broken and the phonon density of states is activated in the
IR spectra. In first approximation newly activated modes are mainly the phonons with the
modulation wavevector \textbf{q$_{m}$}.\cite{petzelt81} One can see that the finger-print
of the modulated structure is the activation of two phonons between 200 and 270\cm, which
is clearly seen in single crystal and the B ceramics. Also the splitting of TO3 modes is
resolved in both samples at low temperatures. TO1 and TO4 modes are split in all samples
already due to tetragonal distortion. We note that IR spectroscopy is very sensitive on
small local breaking of symmetry, which is sometimes hardly resolved in X-ray and
electron diffraction experiments.

\begin{widetext}
\begin{table}[h]
\centering{}
\begin{tabular}{||cc||cc|cc||cc|cc|cc||cc||}
\hline
\multicolumn{2}{||c||}{Mode} &\multicolumn{2}{|c|}{A ceramics} & \multicolumn{2}{c||}{$I4/mcm$ structure} & \multicolumn{2}{c|}{B ceramics}&\multicolumn{2}{c|}{$Imma$ structure}&\multicolumn{2}{c||}{Single crystal}&\multicolumn{2}{c||}{$Pm\overline{3}m$ structure}\\
\multicolumn{2}{||c||}{assignment} &\multicolumn{2}{|c|}{experiment} & \multicolumn{2}{c||}{theory}  & \multicolumn{2}{c|}{experiment}&\multicolumn{2}{c|}{theory}   &\multicolumn{2}{c||}{experiment}& \multicolumn{2}{c||}{theory} \\
\hline\hline
 & & $\omega_{TO}$ & $\Omega_{P}$ & $\omega_{TO}$ & $\Omega_{P}$ & $\omega_{TO}$ & $\Omega_{P}$ & $\omega_{TO}$ & $\Omega_{P}$  & $\omega_{TO}$ & $\Omega_{P}$ & $\omega_{TO}$ & $\Omega_{P}$\\
\hline
 & TO1 & 63 & 1162& 107 & 1323  & 82  & 1290& 98  & 1318 & 81 & 742  & 67 & 1289\\
 & TO1 & 78 & 871 & 128 & 1607  & 107 & 713 & 110 & 1233 & 92 & 1211 &    &     \\
 &     &    &     &     &       &     &     & 131 & 1596 &    &      &    &     \\
 & TO2 & 153& 329 & 154 & 898   & 154 & 473 & 154 &  822 & 156& 470  & 155& 925 \\
 & TO2 &    &     & 156 & 314   &     &     & 155 &  201 &    &      &    &     \\
 &     &    &     &     &       &     &     & 159 &  945 &    &      &    &     \\
 &     &    &     &     &       & 220 & 82  & 237 &  134 & 200& 123  &    &     \\
 &     &    &     & 251 & 40    & 277 & 240 & 283 &  122 & 249& 374  &    &     \\
 & TO3 &    &     &     &       & 433 & 130 & 416 &  224 & 432&  61  &    &     \\
 & TO3 & 431& 348 & 419 & 250   & 442 &  99 & 418 &  230 & 443&  42  &    &     \\
 &     &    &     &     &       &     &     & 516 & 730  &    &      &    &     \\
 & TO4 & 537& 591 & 523 & 732   & 546 & 672 & 523 & 736  & 541& 617  & 537& 824 \\
 & TO4 & 570& 228 & 531 & 718   & 600 &  63 & 537 & 725  & 596& 103  &    &     \\
\hline
\end{tabular}
\caption{List of phonon $\omega_{TO}$ and plasma
$\Omega_{P}=\sqrt{\Delta\varepsilon}\omega_{TO}$ frequencies in EuTiO$_{3}$ ceramics and
single crystal obtained from the fits of IR spectra at 10 K. Experimental parameters are
compared with parameters of polar phonons obtained from first principles calculations\cite{rushchanskii12} in
different crystal structures. All the parameters are in \cm.} \label{phonon-freq}
\end{table}
\end{widetext}

Rushchanskii et al.\cite{rushchanskii12} performed the theoretical analysis of all
possible structures in EuTiO$_{3}$ and came to the conclusion that this material has
three different possible ground states with very similar energies. Our structural and
infrared investigations confirmed a tetragonal structure in the A ceramics prepared by
the conventional method. Single crystals exhibit tetragonal distortion and moreover an
incommensurate modulation. Small crystallites and large internal microstrain in the B
ceramics prepared using high-pressure high-temperature sintering did not allow to resolve
the tetragonal or incommensurate structure, but according to IR spectra, the structure is
the same as in the single crystals. Different structural, infrared, dielectric and
magnetic properties of ceramics and single crystals as well as spread of published phase
transition temperatures to tetragonal phase give evidence for a high sensitivity of
physical and structural properties of EuTiO$_{3}$ on concentration of defects in the
samples.

\section{Conclusion}

Until recently it was assumed that EuTiO$_{3}$ crystallizes in a stable cubic
$Pm\bar{3}m$ structure. Our XRD, electron diffraction as well as Young's modulus and
thermal dilatation studies of conventionally prepared EuTiO$_{3}$ ceramics (marked as A
ceramics) reveal the antiferrodistortive phase transition to tetragonal $I4/mcm$ phase
already near 300\K. It is formed by an antiphase tilting of oxygen octahedra along the
\textbf{c} axis ($a^{0}a^{0}c^{-}$ in Glazer notation). This type of the phase
transition, which is the same as in SrTiO$_{3}$, was very recently confirmed also by
other authors, but at different temperatures from 160 to 282
K.\cite{allieta12,kim12,kohler12} Careful XRD measurements of single crystal as well as
grinded crystal did not reveal the tetragonal distortion down to 100\K. On other hand,
electron diffraction of the same grinded crystal, performed at room temperature just
after cooling down to 100\K, revealed not only tetragonal but also incommensurate
structure, which disappeared after several days. This fact was explained by
incommensurate phase transition at 285\K\, observed very recently by Kim et
al.\cite{kim12}, because this first order phase transition can exhibit a temperature
hysteresis. We propose to explain the observed discrepancies in structural, infrared,
dielectric and magnetic behavior of ceramics and single crystals as well as various
reported critical temperatures by various concentration of Eu$^{3+}$ defects and oxygen
vacancies. Their determination using positron annihilation and M\"{o}ssbauer spectroscopy
are currently in progress.

\begin{acknowledgments}

This work was supported by the Czech Science Foundation (Project No. P204/12/1163),
M\v{S}MT (COST MP0904 projects LD12026 and LD11035), Project Praemium Academiae of ASCR
and the Austrian Science Found (FWF Project No. P23982-N20).

\end{acknowledgments}

\end{document}